\newcommand{\version}{October 20, 2020}
\documentclass[letterpaper,oneside,11pt,pdftex]{article}
\pdfoutput=1
\usepackage[bindingoffset=0.0cm,textheight=22.60cm,hdivide={*,16.1cm,*}, vdivide={*,22.60cm,*}]{geometry}
\usepackage{amsmath,amsfonts,amssymb}
\usepackage{mathtools}
\usepackage[pdftex]{graphicx}
\usepackage[margin=1cm,font=small]{caption}
\usepackage{subcaption}
\usepackage[T1]{fontenc}
\usepackage[utf8]{inputenc}
\usepackage{fouriernc}
\newcommand{\id}{\mathbb{1}}
\usepackage[numbers,square,comma,sort&compress]{natbib}

\usepackage[pdftex,hyperref,svgnames]{xcolor}
\usepackage[pdftex,bookmarksnumbered=true,breaklinks=true,%
colorlinks=true,linktocpage=true,linkcolor=MediumBlue,citecolor=ForestGreen,urlcolor=DarkRed]{hyperref}

\makeatletter
\AtBeginDocument{
  \hypersetup{
    pdfkeywords = {\keywords},
    pdftitle = {\@title},
    pdfauthor = {\@author}
  }
}
\makeatother

\makeatletter

 \@addtoreset{equation}{section}
 \makeatother

\renewcommand{\d}{\delta}

\newcommand{\vth}{\vartheta}

\newcommand{\pa}{\partial}

\newcommand{\nn}{\nonumber}
\newcommand{\eqnref}[1]{Eq. \eqref{#1}}

\newcommand{\ct}{c_{\textrm{T}}}

\newcommand{\gt}{\gamma_{\textrm{T}}}

\newcommand*{\mat}[1]{\mathrm{#1}}

\newcommand{\txt}[1]{\textrm{#1}}

\newcommand{\ftntmark}[1]{\texorpdfstring{\footnotemark[#1]~}{}}

\DeclarePairedDelimiter\abs{\lvert}{\rvert}

\title{\texorpdfstring{\begin{flushright}
        {\small LA-UR-20-25514}
       \end{flushright}\vspace{2em}}{}%
        Clarifying the definition of ‘transonic’ screw dislocations}

\author{Daniel N. Blaschke\ftntmark{1}, Jie Chen\ftntmark{1}, Saryu Fensin\ftntmark{1}, and Benjamin A. Szajewski\ftntmark{2}}

\date{\version}

\newcommand{\keywords}{dislocations in crystals, transsonic motion, limiting velocity}

\graphicspath{{./}{./figures/}}

\begin{document}

 \maketitle

 \begin{center}
 \vspace{-0.3cm}
\renewcommand{\thefootnote}{\fnsymbol{footnote}}
\footnotemark[1] Los Alamos National Laboratory, Los Alamos, NM, 87545, USA\\
\footnotemark[2] CCDC Army Research Laboratory, Aberdeen Proving Ground, MD, 21005, USA
\\[0.5cm]
\ttfamily{E-mail: dblaschke@lanl.gov, jie.chen@lanl.gov, saryuj@lanl.gov, benjamin.a.szajewski.civ@mail.mil}
 \end{center}

\vspace{1.5em}

\begin{abstract}
A number of recent Molecular Dynamics (MD) simulations have demonstrated that screw dislocations in face centered cubic (fcc) metals can achieve stable steady state motion above the lowest shear wave speed ($v_\txt{shear}$) which is parallel to their direction of motion (often referred to as transonic motion).
This is in direct contrast to classical continuum analyses which predict a divergence in the elastic energy of the host material at a crystal geometry dependent `critical' velocity $v_\txt{crit}$.
Within this work, we first demonstrate through analytic analyses that the elastic energy of the host material diverges at a dislocation velocity ($v_\txt{crit}$) which is greater than $v_\txt{shear}$, i.e. $v_\txt{crit} > v_\txt{shear}$.
We argue that it is this latter derived velocity ($v_\txt{crit}$) which separates `subsonic' and `supersonic' regimes of dislocation motion in the analytic solution.

In addition to our analyses, we also present a comprehensive suite of MD simulation results of steady state screw dislocation motion for a range of stresses and several cubic metals at both cryogenic and room temperatures.
At room temperature, both our independent MD simulations and the earlier works find stable screw dislocation motion only below our derived $v_\txt{crit}$. 
Nonetheless, in real-world polycrystalline materials $v_\txt{crit}$ cannot be interpreted as a hard limit for subsonic dislocation motion.
In fact, at very low temperatures our MD simulations of Cu at 10 Kelvin confirm a recent claim in the literature that true `supersonic' screw dislocations with dislocation velocities $v>v_\txt{crit}$ are possible at very low temperatures.
\end{abstract}

\newpage
\tableofcontents

\section{Introduction and background}
\label{sec:intro}

Plasticity in metals is mediated by the motion of linear crystalline defects known as dislocations.
In regimes of high-rate loading~\cite{Clifton:1983}, such as impact~\cite{Clifton:2000,Murr:1996,Austin:2011,Luscher:2016,Gurrutxaga:2016}, the question of how fast dislocations can flow becomes increasingly important towards understanding how high rates of material deformation are accommodated;
see also \cite{Gurrutxaga:2020} for a recent nice review of high speed dislocation dynamics.
In particular, the strain rate of a material is proportional to the product of (mobile) dislocation density and the average dislocation velocity via Orowan's relation \cite{Blaschke:2019a}.
As such, the question of whether dislocations can reach supersonic speeds directly influences our understanding of strain rate as a function of stress:
If a fraction of dislocations move supersonically, the average dislocation speed in Orowan's relation is higher and this in turn implies that lower mobile dislocation densities are necessary to explain observed strain rates.
Thus, our understanding of dislocation mobility is also interconnected with our understanding of dislocation density evolution.

A suite of recent Molecular Dynamics (MD) simulations~\cite{Olmsted:2005,Marian:2006,Daphalapurkar:2014,Tsuzuki:2008,Tsuzuki:2009,Oren:2017,Ruestes:2015,Gumbsch:1999,Li:2002,Jin:2008,Peng:2019} as well as some experimental data~\cite{Nosenko:2007} suggest that
dislocations can reach transonic or even supersonic velocities.
Most of these results pertain to edge dislocations in fcc metals like Al, Cu, and Ni \cite{Olmsted:2005,Marian:2006,Daphalapurkar:2014,Tsuzuki:2008,Tsuzuki:2009,Oren:2017} and bcc metals\footnote{
We focus here on MD results that study the question whether supersonic dislocation motion can be achieved.
For example, Ref. \cite{Gilbert:2011} studies screw dislocation motion in bcc $\alpha$-Fe, but only below 1 km/s.
}
like W and Ta \cite{Gumbsch:1999,Li:2002,Jin:2008,Ruestes:2015}.
Less is known about screw dislocations.
As an example, screw dislocation motion has been studied within fcc Al in\footnote{
Ref. \cite{Cho:2017} also study subsonic dislocation motion in Al, putting their emphasis on dislocation character dependence, but do not comment on what happens beyond stresses of 1GPa (the highest they study).}
\cite{Vandersall:2004,Olmsted:2005}, and the authors report that twinning is activated at higher stresses preventing stable dislocation motion above the lowest shear wave speed of the material in the direction of dislocation motion.
Other authors report that stable transonic screw dislocation motion in fcc Cu \cite{Oren:2017} and Ni \cite{Olmsted:2005,Marian:2006} is possible and that twinning preventing stable dislocation motion occurs only at even higher speeds referred to as the `deep transonic regime'.
Recently, Ref. \cite{Peng:2019} reported screw dislocation motion in fcc Cu up to 3.5 km/s in simulations undertaken at 1 Kelvin.
Furthermore, while edge dislocation speeds seem to saturate below the lowest shear wave speed before discontinuously shifting into the transonic regime above some critical stress, the reported screw dislocations in Cu \cite{Oren:2017} and Ni \cite{Olmsted:2005,Marian:2006}  enter into a  purported transonic regime in a much smoother fashion adding to the mystery as to why their behavior is so different not only from Al but also from edge dislocations.

The attentive reader will have noticed by now that the term ``transonic'' is nebulous, except in isotropic materials where it refers to velocities between well-defined transverse and longitudinal sound speeds.
In general anisotropic crystals, sound speeds are direction dependent and indeed most of the references cited above define ``transonic'' as above the lowest shear wave speed pertaining to the direction of dislocation motion but below a (direction dependent) longitudinal sound speed.
In fact, the first such MD simulations were carried out for W \cite{Gumbsch:1999} which has a Zener anisotropy ratio near unity and is therefore considered to be an almost ``isotropic'' bcc metal.
Since W is effectively ``isotropic'' these simulations do not show any dependence on crystalline anisotropy.
Some authors compare their dislocation velocity results to the average isotropic transverse sound speed \cite{Vandersall:2004,Wang:2008}, whose relevance for dislocation dynamics in anisotropic crystals is questionable.

Our objective in the present work is to reexamine the alleged transonic screw dislocations in Cu and Ni at both room temperature and at very low temperatures.
We demonstrate below that the lowest shear wave speed has no relevance for any dislocation motion:
Taking elastic anisotropy into account, limiting ``critical'' or ``forbidden'' dislocation velocities must be calculated specifically for a character dependent dislocation field and are generally different from any sound speed.
Despite this fact, depending on the crystal and slip system geometry the lowest shear wave speed may happen to be close to or even coincide with the dislocation's critical velocity  (e.g. the case for edge dislocations in fcc) \cite{Blaschke:2017lten}.
This is likely the reason that this point has been overlooked in the earlier literature.
As we demonstrate below, these two velocities are noticeably different for fcc screw dislocations.
Our predicted differences may help to elucidate various discrepancies found within the earlier literature (cf. refs. \cite{Olmsted:2005,Marian:2006,Oren:2017,Peng:2019}).
In fact, the earlier results of Refs. \cite{Olmsted:2005,Marian:2006,Oren:2017} show stable motion for screw dislocations in Cu and Ni only at velocities which are \emph{slower} than their critical velocity.
At room temperature, our present simulations support this conclusion for Cu, but at 10 Kelvin we
confirm the results of Ref. \cite{Peng:2019} that supersonic screw dislocations in Cu (i.e. with velocity $v>v_\txt{crit}$) are possible at very low temperatures.

\section{Speeds of sound and limiting velocities of dislocations}
\label{sec:steadystate_anis}

Both sound waves and moving dislocations may be quantified in terms of spatiotemporal displacement fields applied to atoms with respect to a perfect lattice configuration.
For sound waves, these displacement fields are generally small and smooth; this is not the case for line defects (dislocations).
Independent of motion, dislocations are most readily described by a  non-vanishing Burgers vector which introduces a spatial displacement discontinuity across the slip plane.
In the continuum limit, the displacement field $u_i$ satisfies the balance of linear momentum and the leading order stress-strain relations known as Hooke's law:
\begin{align}
\sigma_{ij,i}&=\rho\ddot u_j\,, &
 \sigma_{ij}&= 
C_{ijkl}\epsilon_{kl}=
C_{ijkl} u_{k,l}
\,, & \nonumber\\
 \epsilon_{kl}&\equiv(u_{k,l}+u_{l,k})/2
 \,. \label{eq:Hooke}
\end{align}
Within these equations, $\sigma_{ij}$ denotes stress, $\epsilon_{kl}$ is the infinitesimal strain tensor (i.e. the symmetrized displacement gradient field), $\rho$ the material density, and we have introduced the common shorthand notation $u_{k,l}\equiv\pa_l u_k$ for partial derivatives.
$C_{ijkl}$ denotes the components of the fourth rank tensor $\mat{C}$ of second order elastic constants (SOEC), also known as the 'stiffness tensor'.
Within these expressions, we employ index notation and Einstein's summation convention where the sum over repeated indices is implied.
In addition, a dot denotes differentiation with respect to time.

In a general anisotropic crystal, the sound speeds in the direction of motion are determined from \cite{Bacon:1980}
\begin{align}
 \det\left(\hat{v}\cdot\mat{C}\cdot\hat{v}-\rho v^2\id\right)\Big|_{v=v_{\txt{shear}}}=0
 \,, \label{eq:sound1}
\end{align}
with $\hat{v}=\vec{v}/v$ denoting the unit vector in the direction of wave propagation.
The three solutions for sound speed $v$ in the direction of $\hat{v}$, may or may not coincide with a pure `shear' wave.
In the isotropic limit, all directions $\hat{v}$ yield the same three solutions, albeit the two lower ones coincide and are the well-known transverse sound speed $\ct$.
We show solutions for $v_\txt{shear}$ for several fcc metals within Table \ref{tab:values-metals} .

Within this work, we focus on cubic symmetry (e.g. fcc or bcc), and therefore the tensor of SOECs within the crystal reference frame is
\begin{align}
 C_{ijkl}&=c_{12}\d_{ij}\d_{kl}+c_{44}\left(\d_{ik}\d_{jl}+\d_{il}\d_{jk}\right)-H\sum_{\alpha=1}^3\d_{i\alpha}\d_{j\alpha}\d_{k\alpha}\d_{l\alpha}
 \,,\label{eq:C2_cubic}
\end{align}
where $H\equiv 2(c_{44}-c')$ and $c'=(c_{11}-c_{12})/2$.
The first two terms are invariant under rotation.
The third term (which includes the explicit summation over $\alpha$) however is reference frame dependent as it explicitly depends on the crystal basis vectors;
within a Cartesian coordinate system the crystal basis vectors coincide with $\hat x_i=\d^1_i$, $\hat y_i=\d^2_i$, and $\hat z_i=\d^3_i$.
As such, the third term breaks isotropy for $H\neq0$ (respectively $c'\neq c_{44}$).
Within cubic crystals, dislocation line directions are not parallel to crystal basis vectors, and as such it is convenient to rotate the balance of linear momentum equation~\eqref{eq:Hooke} into a coordinate system with a basis vector coincident with the dislocation line direction.

A pure screw dislocation is comprised of a Burgers vector $\vec{b}$ parallel to the line sense unit vector ($\hat{t}$).
Choosing coordinates such that $\hat{z}=\hat{t}$, such a \emph{pure} screw dislocation can only be stable, if its displacement field has the form $u_i=(0,0,u_z(x,y,t))$
and if the stress-strain relations \eqref{eq:Hooke} yield vanishing stress in the $x$-$y$ plane, i.e. $\sigma_{xx}=\sigma_{yy}=\sigma_{xy}=0$.
As discussed in Ref. \cite[Chapter 13]{Hirth:1982}, this is the case for slip systems where the $x$, $y$ plane (in coordinates aligned with the dislocation) is a reflection plane, and we will show below that this is the case for the 12 fcc slip systems.

In order to solve the differential equation \eqref{eq:Hooke} for the displacement in coordinates aligned with the dislocation, we must rotate the tensor of SOEC accordingly, i.e.
\begin{align}
C'_{ijkl} = U_{ii'} U_{jj'} U_{kk'} U_{ll'}C_{i'j'k'l'} 
\end{align}
where $U_{ij}$ denotes the rotation matrix which in this work rotates from the crystal reference frame into a dislocation oriented reference frame.
In addition to $\hat{z}$ along $\hat{t}$, we also select a second basis vector $\hat{y}$ aligned with slip plane normal $\hat{n}_0$.
Within the dislocation oriented reference frame, the differential equation governing a pure screw dislocation reduces to
\begin{align}
\rho\partial_t^2u_z(x,y,t)=(A\partial_x^2+B\partial_x\partial_y+D\partial_y^2)u_z(x,y,t)
\,, \label{eq:diffeq_screw_gen}
\end{align}
where
\begin{align}
A =& C'_{1331} \,, &
B =& C'_{1332} + C'_{2331}\,, &
D =& C'_{2332}
\,. \label{eq:Coefficients}
\end{align}
The coefficients $A$, $B$, and $D$ are linear combinations of the SOEC and also depend on the rotation matrix $U$.
In the isotropic limit, $A = D=c_{44}$, $B = 0$, and the differential operator on the right hand side of \eqref{eq:diffeq_screw_gen} reduces to the 2-dimensional Laplacian operator, i.e. $c_{44}\nabla^2 u_z(x,y,t)$.
The other two components of the balance of linear momentum, $\partial_i\sigma_{ix}=0=\partial_i\sigma_{iy}$ identically, if the slip system has the required symmetry properties for a pure screw dislocation as detailed above.
The boundary conditions appropriate for a screw dislocation starting at the origin with constant velocity with Burgers vector $b\hat{z}$ is \cite{Markenscoff:1980}
\begin{align}
\lim_{\eta\to0^+}\left(u_z(x,\eta,t) - u_z(x,-\eta,t)\right) &= b\Theta(x-vt)
\,, \label{eq:bc_iso}
\end{align}
where $\Theta$ denotes the Heaviside function.
Boundary condition \eqref{eq:bc_iso} enforces a slip discontinuity across the slip plane, i.e. the displacement is shifted by $b$ along $\hat{z}$ when crossing the $y = 0$ plane.

Since the dislocation moves at constant velocity $v$, the time variable may be eliminated entirely from the differential equation via substitution of $x'=x-vt$.
Specifically, a coordinate system which translates with the dislocation is adopted with displacement field $u_z(x',y)$.
Within this coordinate system, application of the chain rule to partial derivatives yields $\partial_t = -v\partial_x$ and $\partial_x=\partial_{x'}$, and therefore \eqref{eq:diffeq_screw_gen} takes the form
\begin{align}
0&=\left[A\partial_{x'}^2+B\partial_{x'}\partial_y+D\partial_y^2 - {\rho}{v^2}\partial_{x'}^2\right]u_z(x',y)
\nn\\
&=\left[\left(A-\frac{B^2}{4D} - {\rho}{v^2}\right)\partial_{x'}^2+D\left(\frac{B}{2D}\partial_{x'}+\partial_y\right)^2\right]u_z(x',y)
\,.
\end{align}
The solution (consistent with the boundary conditions) is readily verified to be
\begin{align}
u_z(x',y)
&= \frac{b}{2\pi} \arctan\left(\frac{y\sqrt{\frac{1}{{D}} \left(A -  \frac{{B}^2}{4{D}} - \rho{v^2} \right)}}{-x'+\frac{{B}}{2{D}}y}\right)
\,, \label{eq:uz_steadystate_anis}
\end{align}
which represents a closed form solution for the displacement field of a steady state moving screw dislocation within an anisotropic crystal.
Eq. \eqref{eq:uz_steadystate_anis} exhibits a limiting velocity $v_\txt{crit}=\sqrt{\frac{1}{\rho}\left(A-\frac{{B}^2}{4{D}}\right)}$ since its gradient (and consequently the strain field and its elastic self-energy) diverges at this velocity:
\begin{align}
\partial_{x'} u_z(x',y) &
=\frac{b}{2\pi}\frac{y}{\gamma_\txt{crit}\left(x'-\frac{{B}}{2{D}}y\right)^2   + {y^2}/{\gamma_\txt{crit}}}
\,,\qquad\qquad
\partial_y u_z(x',y) 
= \frac{b}{2\pi} \frac{ - x' }{\gamma_\txt{crit}\left(x'-\frac{{B}}{2{D}}y\right)^2   + y^2/\gamma_\txt{crit}}
\,,\nn\\
\gamma_\txt{crit} & 
= \frac{1}{\sqrt{\frac{1}{{D}} \left(A -  \frac{{B}^2}{4{D}} - \rho{v^2} \right)}}
= \frac{1}{\sqrt{\frac{1}{{D}} \left(A -  \frac{{B}^2}{4{D}}\right)\left(1 - \frac{v^2}{v_\txt{crit}^2} \right)}}
\,. \label{eq:analyticsteadystate}
\end{align}
Along the lines $x'=x-vt=\frac{{B}}{2{D}}y$, both terms tend to $\sim 1/\sqrt{1-v^2/v_\txt{crit}^2}$ as $v$ approaches the critical velocity (see Fig. \ref{fig:divergingsteadystate}) and this divergence propagates to the screw dislocation's strain self-energy.

In the isotropic limit ($\mu\equiv A=D$, $B=0$, and $v_\txt{crit}=\ct=\sqrt{\mu/\rho}$), $u_z$ simplifies to Eshelby's well known steady-state solution \cite{Eshelby:1949} for screw dislocations
\begin{align}
u_z^\txt{iso}(x',y) &
= \frac{b}{2\pi} \arctan\left(\frac{y}{-\gt x'}\right)
\,, \qquad\qquad
\gt  = \frac{1}{\sqrt{\left(1 - \frac{v^2}{\ct^2} \right)}}
\,.
\end{align}
Note that we have placed the minus sign next to $x'$ inside the argument of the arc-tangent function to fulfill our boundary condition \eqref{eq:bc_iso}, since $\arctan()$ is a multi-valued function and its result at $y=0$ is either $0$ or $\pi$ depending on the quadrant from which the $y\to0$ limit is taken.

We emphasize that \eqref{eq:uz_steadystate_anis} is applicable to any crystal geometry, provided the slip plane is a reflection plane \cite[Chapter 13]{Hirth:1982}.
In cubic crystals this is the case for all 12 fcc slip systems but none of the 48 bcc slip systems.
Likewise, hcp, tetragonal, and other crystal geometries exhibit a number of slip planes which fulfill this symmetry requirement.
We further direct our attention to fcc crystals, because it is there that we wish to clarify previous claims in the literature of observing supersonic screw dislocations.
We do this by making very clear the distinction between the critical velocity and the shear wave speed, and by conducting additional MD simulations of our own.
Furthermore, we emphasize that the solution \eqref{eq:analyticsteadystate} coincides (numerically) with the well-known one derived with the so-called `integral method' \cite{Bacon:1980}, and we verify this explicitly for fcc metals within Appendix \ref{sec:comparison}.
The reason we present \eqnref{eq:analyticsteadystate} here is merely to make the point that $v_\txt{crit}\neq v_\txt{shear}$ clear because the value of $v_\txt{crit}$ is somewhat obscured in the integral method due to a numerical angle integration as discussed in Appendix \ref{sec:comparison}; we do not claim to have found a new theory.

\subsection*{FCC slip systems}

\begin{figure}[ht]
\centering
 \includegraphics[width=0.5\textwidth]{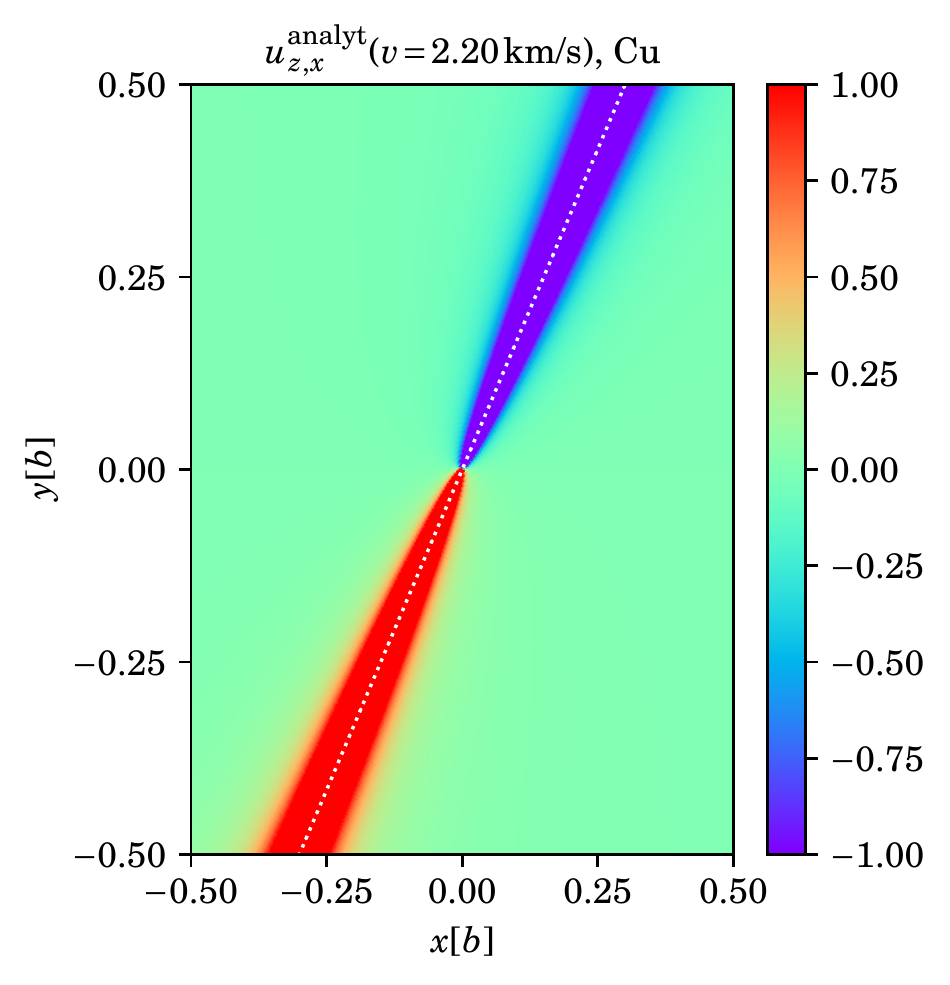}%
 \includegraphics[width=0.5\textwidth]{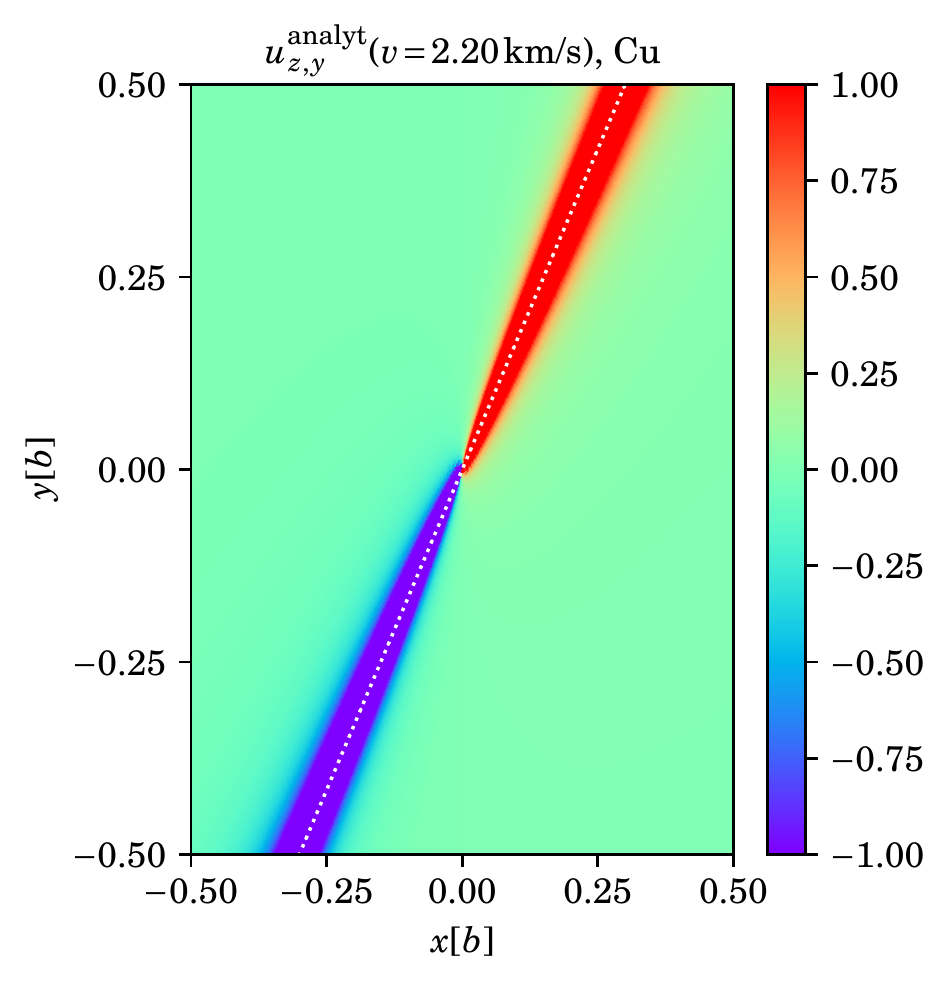}
\caption{We show $u_{z,x}$, $u_{z,y}$ (see \eqref{eq:analyticsteadystate_fcc}) for Cu at a dislocation velocity $v=0.999 v_\txt{crit}$ in the positive $x$ direction.
The dotted lines highlight the contours $x'=y\frac{B}{2D}=y\frac{\sqrt{2}(c_{44}-c')}{(c_{44}+2c')}$ along which the solution diverges at $v=v_\txt{crit}$.}
\label{fig:divergingsteadystate}
\end{figure}

We now determine the rotation matrix $U$ and coefficients $A$, $B$, $D$ for (one of) the 12 fcc slip systems explicitly.
Characteristic of fcc metals is a Burgers unit vector $\hat{b}=(1,1,0)/\sqrt{2}$ and slip plane normal $\hat{n}_0=(1,-1,1)/\sqrt{3}$ in Cartesian coordinates.
For a screw dislocation, the line sense is parallel (or antiparallel) to $\hat{b}$, and following \cite{Blaschke:2017lten} $\hat{t}(\vth) = \frac1b\left[\vec{b}\cos\vth+\vec{b}\times\hat{n}_0\sin\vth\right]$ with character angle $\vth=0$ we presently have $\hat{t}(0) = \hat{b}=\frac1b\vec{b}$ for a screw dislocation.
Assuming a straight dislocation that is much longer than its Burgers vector length, the only velocity component that matters is the one perpendicular to the dislocation line contained in the slip plane, i.e. $\vec{v} = \pm v \hat{v}$ where $\hat{v} = \hat{n}_0\times\hat{t}(0)= \hat{n}_0\times\hat{b} = (-1,1,2)/\sqrt{6}$.
To derive $A$, $B$, and $D$ within \eqref{eq:Coefficients}, we need the rotation matrix that aligns $\hat{t}\|\hat{z}$, $\hat{n}_0\|\hat{y}$, and $\hat{v}\|\hat{x}$.
Defining unit vectors in the crystal reference frame $\hat{e}_1 \equiv (1,0,0)$, $\hat{e}_2 \equiv (0,1,0)$ and $\hat{e}_3 \equiv (0,0,1)$, the rotation matrix that fulfills $\mat{U}\cdot \hat{b} = \hat{z}$, $U\cdot \hat{n}_0 = \hat{y}$, and $U\cdot\hat{v}=\hat{x}$,
follows as $\mat{U} = [\hat{v},\hat{n}_0,\hat{t}]^T \cdot [{\hat{e}_1}^T, {\hat{e}_2}^T, {\hat{e}_3}^T]$, where $^T$ denotes the transpose operator.
For the fcc slip system defined by $\hat{b}$ and $\hat{n}_0$ above, we find
\begin{align}
\mat{U}^{\txt{fcc}} &
=\frac{1}{\sqrt{6}}\left(\begin{array}{ccc}
-1 & 1 & 2 \\
\sqrt{2} & -\sqrt{2} & \sqrt{2} \\
\sqrt{3} & \sqrt{3} & 0
\end{array}\right)
\,. \label{eq:Urot}
\end{align}
Employing  $C'_{ijkl} = U_{ii'} U_{jj'} U_{kk'} U_{ll'}C_{i'j'k'l'} $ it is readily verified that in the dislocation oriented reference frame, the only non-vanishing stress components of an fcc pure screw dislocation are
$\sigma_{13}= \sigma_{31} $ and $\sigma_{23}= \sigma_{32} $.
Since $\sigma_{11}=\sigma_{12}=\sigma_{22}=0$, 
the present slip system fulfills the symmetry requirements allowing us to study pure screw dislocations.
The divergence of this stress tensor straightforwardly computes to $\partial_i\sigma_{ij} = \left(0,0,A \partial_x^2u_{z} +B\partial_x\partial_y u_z + D\partial_y^2u_z\right)$,
see \eqref{eq:Coefficients},
with coefficients
\begin{align}
A & = \frac13(c'+2c_{44})
 = c_{44}-\frac16H\,, &
B & = \frac{2\sqrt{2}}{3}(c_{44}-c') = \frac{\sqrt{2}}{3}H\,, &
D & = \frac1{3}(c_{44}+2c') = c_{44} - \frac13H
\,.
\end{align}
From these expressions, we see that $A$ and $D$ are differently weighted average shear moduli, $B$ is proportional to the difference of the largest and smallest shear modulus $H$, a measure of anisotropy.
As such, $A$ and $D$ are always positive whereas $B$ is positive for Zener anisotropy ratios $Z_A=c_{44}/c'$ greater than one and negative otherwise.
As expected, in the limit $H \rightarrow 0$ (where $c'\to c_{44}$), $A = D = c_{44}$ and $B = 0$.
One may repeat this exercise for the other 11 fcc slip systems to check that indeed all of them yield the same coefficients above.
The steady-state limiting velocity is therefore
\begin{align}
v_\txt{crit}^\txt{fcc,screw} &=\sqrt{\frac{A-\frac{B^2}{4D}}{\rho}}
=\sqrt{\frac{3c'c_{44}}{\rho(c_{44}+2c')}}
=\sqrt{\frac{c_{44}\left(c_{44} - \frac12H\right)}{\rho\left(c_{44} - \frac13H\right)}}
\,, \label{eq:vcrit_fccscrew}
\end{align}
which, as expected, is always real and positive independent of Zeners ratio.
We have tabulated $v_\txt{crit}^\txt{fcc,screw}$ for several fcc metals within Table \ref{tab:values-metals}.
For comparison, we have also tabulated $v_\txt{shear}$ as detailed in \eqref{eq:sound1}.
The full steady-state solution \eqref{eq:uz_steadystate_anis} for a pure fcc screw dislocation is then
\begin{align}
u_z(x-vt,y)
&=  \frac{b}{2\pi} \arctan\left(\frac{3y\sqrt{c'c_{44} \left(1- \frac{v^2}{(v_\txt{crit}^\txt{fcc})^2} \right)}}{-(c_{44}+2c')(x-vt)+{\sqrt{2}(c_{44}-c')}y}\right)
\,, \label{eq:uz_steadystate_fcc}
\end{align}
whose static limit ($v\to0)$ coincides with the solution derived in Ref. \cite[Eq. (13-128)]{Hirth:1982} upon verifying that $c'_{44}=D$, $c'_{55}=A$, $c'_{45}=B/2$, and $c'_{44}c'_{55}-c'^2_{45}=AD-B^2/4=c'c_{44}$ for our present slip system.
The gradient of $u_z$, \eqref{eq:analyticsteadystate}, presently reads with these coefficients:
\begin{align}
\partial_{x'} u_z(x',y) &
= y \tilde{u}(x',y)
\,, \qquad\qquad
\partial_y u_z(x',y) 
= -x' \tilde{u}(x',y)
\,,\nn\\
\tilde{u}(x',y) &
=\frac{\frac{b}{2\pi}}{\gamma_\txt{crit}\left(x'-\frac{\sqrt{2}(c_{44}-c')}{(c_{44}+2c')}y\right)^2   + {y^2}/{\gamma_\txt{crit}}}
=\frac{\frac{b}{2\pi}}{\gamma_\txt{crit}\left(x'-\frac{H}{\sqrt{2}\left(3c_{44}-H\right)}y\right)^2   + {y^2}/{\gamma_\txt{crit}}}
\,,\nn\\
\gamma_\txt{crit} &
= \frac{(c_{44}+2c')}{3\sqrt{c'c_{44}\left(1 - \frac{v^2}{v_\txt{crit}^2} \right)}}
= \frac{\left(c_{44}-\frac13H\right)}{\sqrt{c_{44}\left(c_{44}-\frac12H\right)\left(1 - \frac{v^2}{v_\txt{crit}^2} \right)}}
\,. \label{eq:analyticsteadystate_fcc}
\end{align}
The divergence at $x'=y\frac{\sqrt{2}(c_{44}-c')}{(c_{44}+2c')}$ for $v\to v_\txt{crit}$ is highlighted in Fig. \ref{fig:divergingsteadystate} at the example of Cu.

We note that the steady-state solution \eqref{eq:uz_steadystate_anis} is general and applies to all crystal symmetries whose slip systems are reflection planes as described above.
In particular, this is also the case for a number of hcp slip systems where $A$, $B$, and $D$ can be determined the same way as above.
None of the 48 bcc slip systems fulfill this requirement, however.

With regard to pure edge dislocations in fcc metals, we note that in principle one must determine the critical velocities from the differential equations as well, but it so happens that in this particular case they coincide with the lowest shear wave speed \cite{Blaschke:2017lten}.
This is best seen from the steady-state solution for dislocations of arbitrary character angle $\vth$ via the well-known 'integral' method, which itself is a generalization of Stroh's method, see~\cite{Bacon:1980}.
The latter solution depends on numerically integrating one angle and thus obscures the value of $v_\txt{crit}$.
In Appendix \ref{sec:comparison} we verify consistency between our present solution and the integral method solution for $\vth=0$.
The integral method can also be used to numerically determine $v_\txt{crit}$ for bcc slip systems (in contrast to \eqref{eq:analyticsteadystate_fcc} above which is only valid for fcc but yields an exact analytic expression for $v_\txt{crit}$ in this case).

\section{MD simulations for aluminum, copper, and tantalum}

\begin{table}[h!t!b]
\centering
\caption{\label{tab:values-metals}We list elastic constants and densities of Al, Cu, and Ta as calculated from the potentials used within our MD simulations at 300K and at 0K.
For comparison, we also list experimentally determined values at ambient conditions, i.e. the values in parenthesis are taken from Refs.~\cite{CRCHandbook,Thomas:1968,Epstein:1965,Leese:1968,Bolef:1961,Soga:1966,Lowrie:1967}.
In the last two columns we show the computed values for the lowest shear wave speed $v_\txt{s}^\txt{screw}$ in the glide direction (indicated in square brackets) and the critical velocity $v_\txt{crit}^\txt{screw}$ for pure screw dislocations.
The lowest shear wave speeds were determined from \eqnref{eq:sound1}, and the critical velocities were computed from the exact expression \eqref{eq:vcrit_fccscrew} for fcc metals and determined numerically from $\det(nn)=0$ within \eqref{eq:ukl-sol} for bcc metals.}
\begin{tabular}{lccccc|cc}
\hline
\hline
\vspace{0.05cm}
	& T[K] &	$\rho$[g/ccm]	&	$c_{11}$[GPa]	&	$c_{12}$[GPa]	& $c_{44}$[GPa]	& $v_\txt{s}$[km/s] & $v_\txt{crit}^\txt{screw}$[km/s] \\
\hline 
fcc	&			&			&		&	&	& &  \\
\hline
Al	& 0 & 2.70	&	116.8	&	60.1	&	31.7	& 3.30\quad[112] & 3.36\\
Al	& 300 &	2.65	&	106.4	&	57.2	&	27.8	& 3.11\quad[112] & 3.17\\
	& &	(2.70)	&	(106.75)	&	(60.41)	&	(28.34)	 & (3.03) & (3.13)\\ 
Cu	&  0 &	8.93	&	169.9	&	122.6	&	76.2	& 2.06\quad[112] & 2.21 \\
Cu	& 300 &	8.81	&	172.5	&	123.6	&	75.3	& 2.08\quad[112] & 2.25 \\
	& &	(8.96)	&	(168.3)	&	(121.2)	&	(75.7) & (2.05) & (2.20) \\ 
Ni	& 300 &	(8.90)	&	(248.1)	&	(154.9)	&	(124.2) & (2.75) &  (3.00) \\ 
\hline
\hline
bcc	&			&			&		&	& & & 	\\
\hline
Ta	& 0 & 16.66	&	266.9	&	160.4	&	86.0 & 1.79\quad[110] & 1.79	\\
Ta	& 300 &	16.60	&	258.4	&	162.5	&	87.4	& 1.70\quad[110] & 1.70	\\
	& &	(16.4)	&	(260.2)	&	(154.4)	&	(82.55)	& (1.80) & (1.80)	\\
	&			&			&		&	& & (1.94)\quad[112]	&	(1.92)\\ 
	&			&			&		&	& & (1.83)\quad[541] & (1.81)	\\
\hline
\hline
\end{tabular}
\end{table}

\begin{figure}[ht]
\centering
\includegraphics[width=\textwidth]{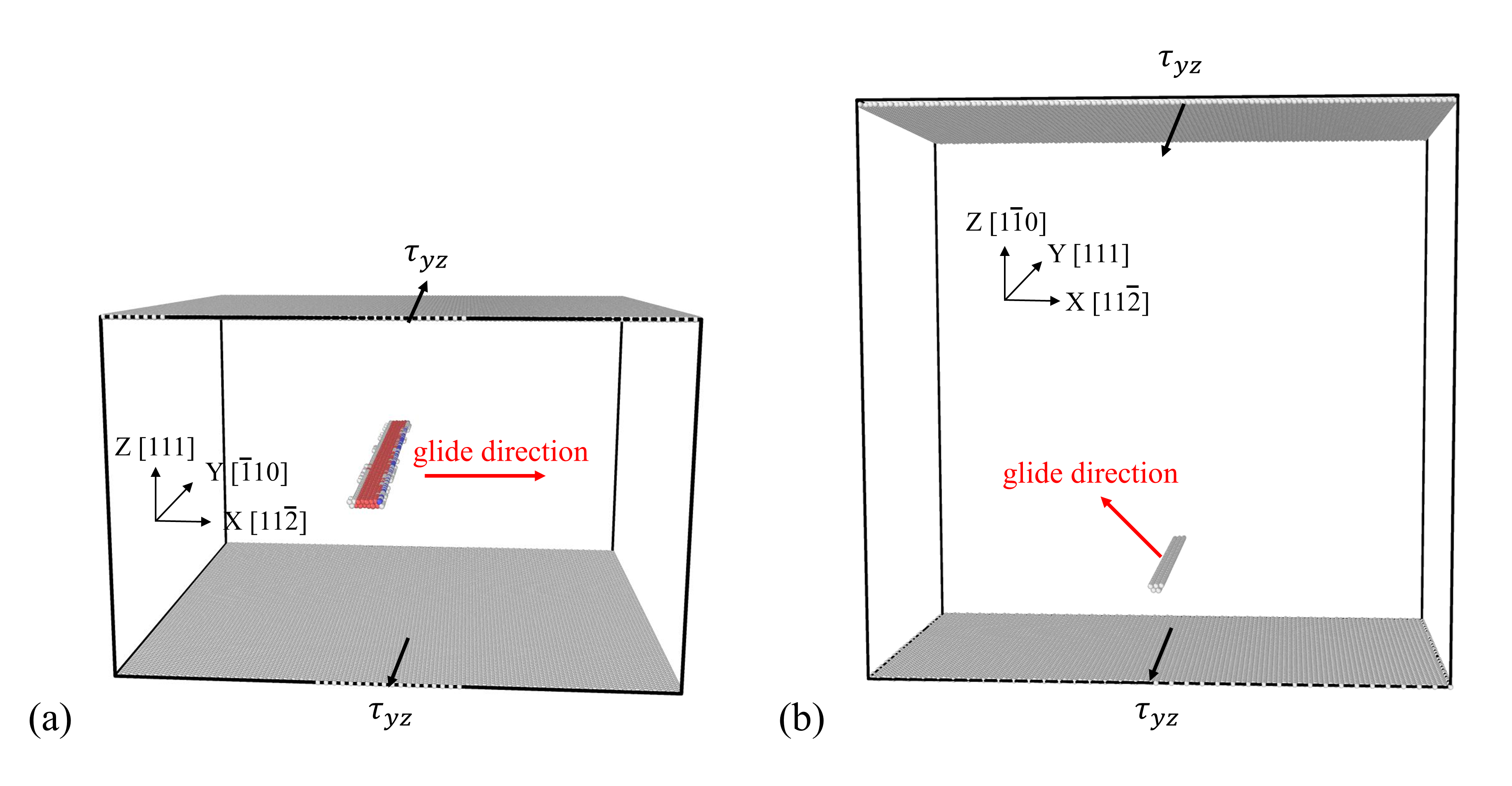}
\caption{System setup for evaluating mobility of screw dislocation in (a) Al or Cu, (b) Ta. For Al or Cu, fcc atoms are removed for visualizing the dislocation line, whereas for Ta, bcc atoms are removed for visualizing the dislocation line.}
\label{fig:setup}
\end{figure}

\begin{figure}[ht]
\centering
\includegraphics[width=\textwidth]{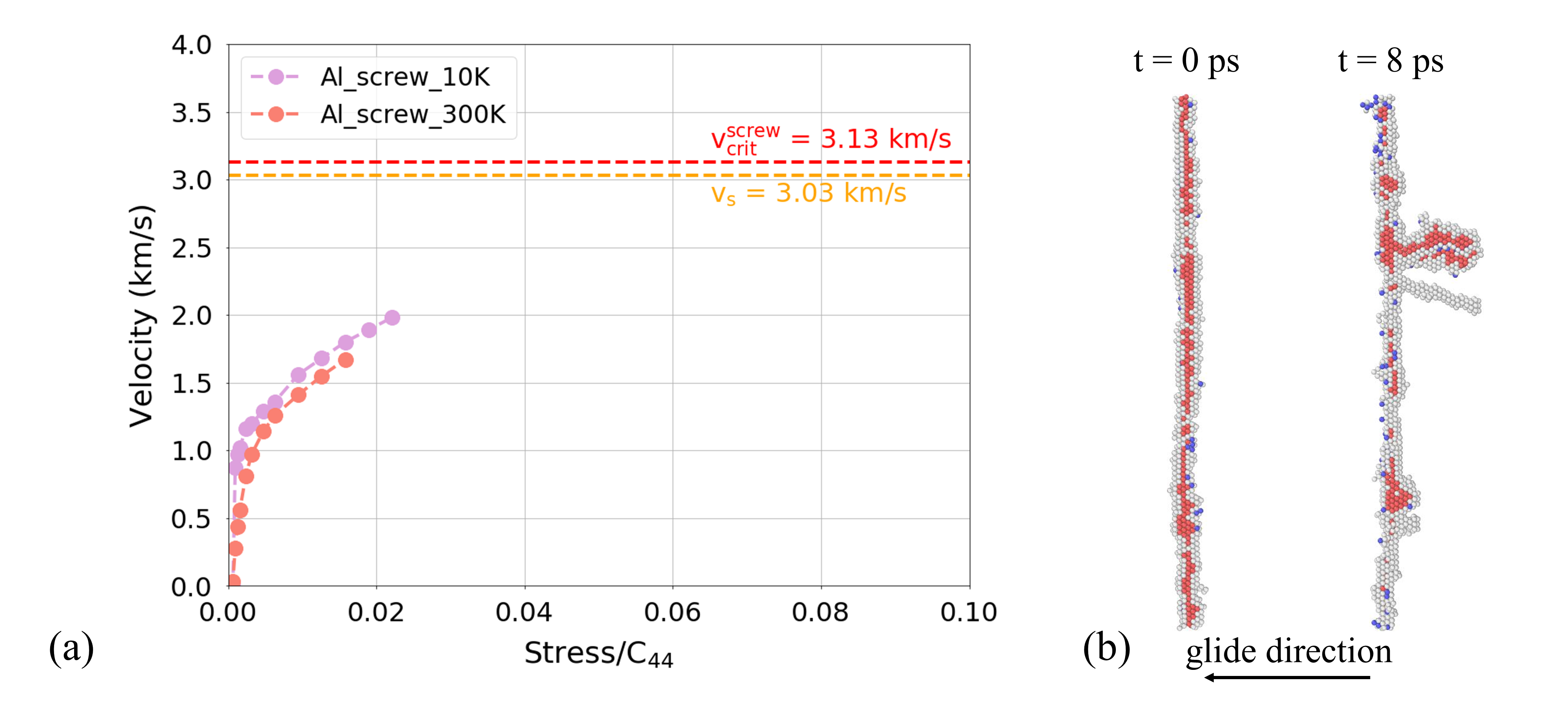}
\caption{(a) Dislocation velocity versus applied stress for screw dislocations in Al ($T = 10$ K, $300$ K), (b) nucleation of extended dislocations from screw dislocations in Al at an applied shear stress of $600$ MPa ($T = 300$ K).}
\label{fig:MD_Al}
\end{figure}

\begin{figure}[h!t]
\centering
\includegraphics[width=\textwidth]{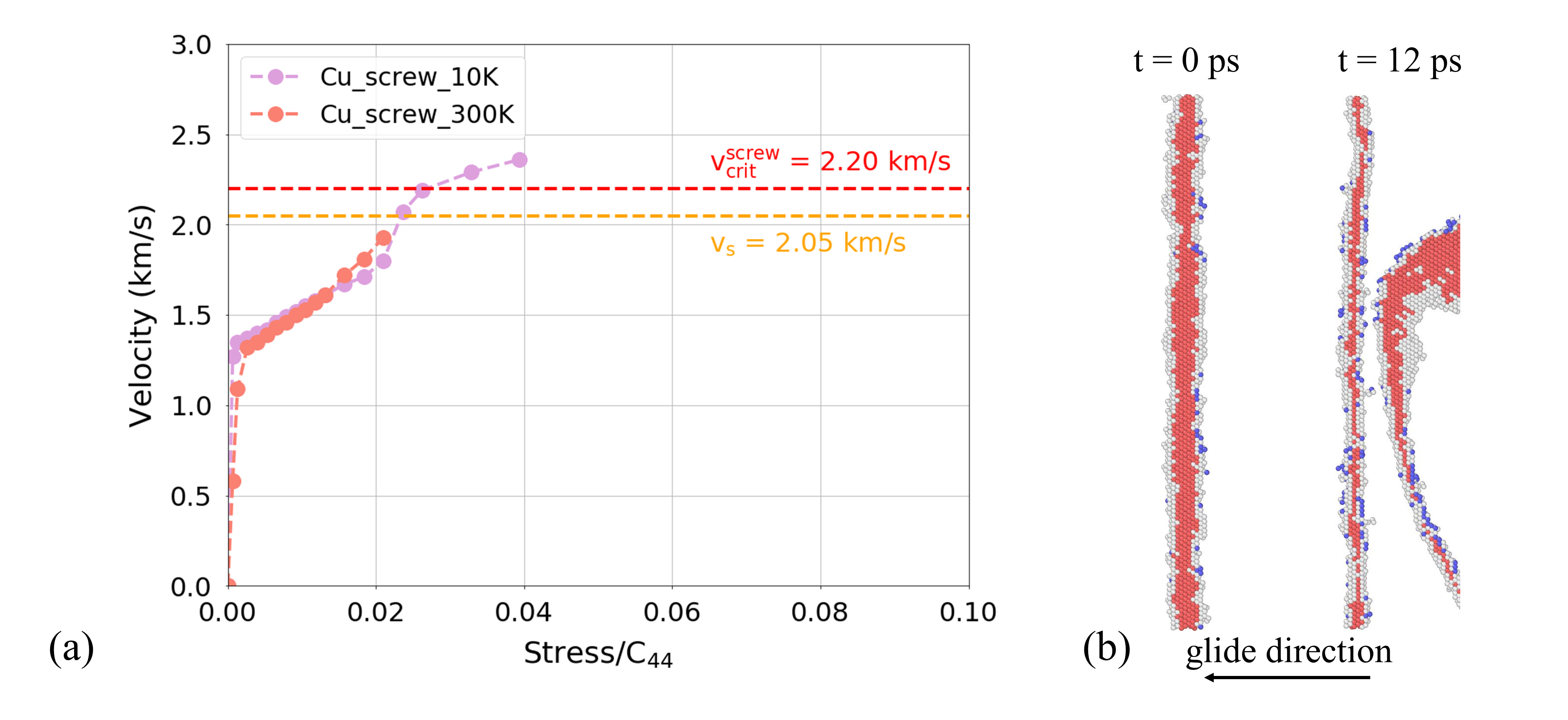}
\caption{We show screw dislocation velocity as a function of applied stress in Cu ($T = 10$ K, $300$ K) resulting from MD simulations as described in the main body of this paper.
In this case, the lowest shear wave speed for Cu is $\sim2.05$ km/s whereas the critical velocity is $\sim2.20$ km/s.}
\label{fig:MD_Cu}
\end{figure}

\begin{figure}[h!t]
\centering
\includegraphics[width=\textwidth]{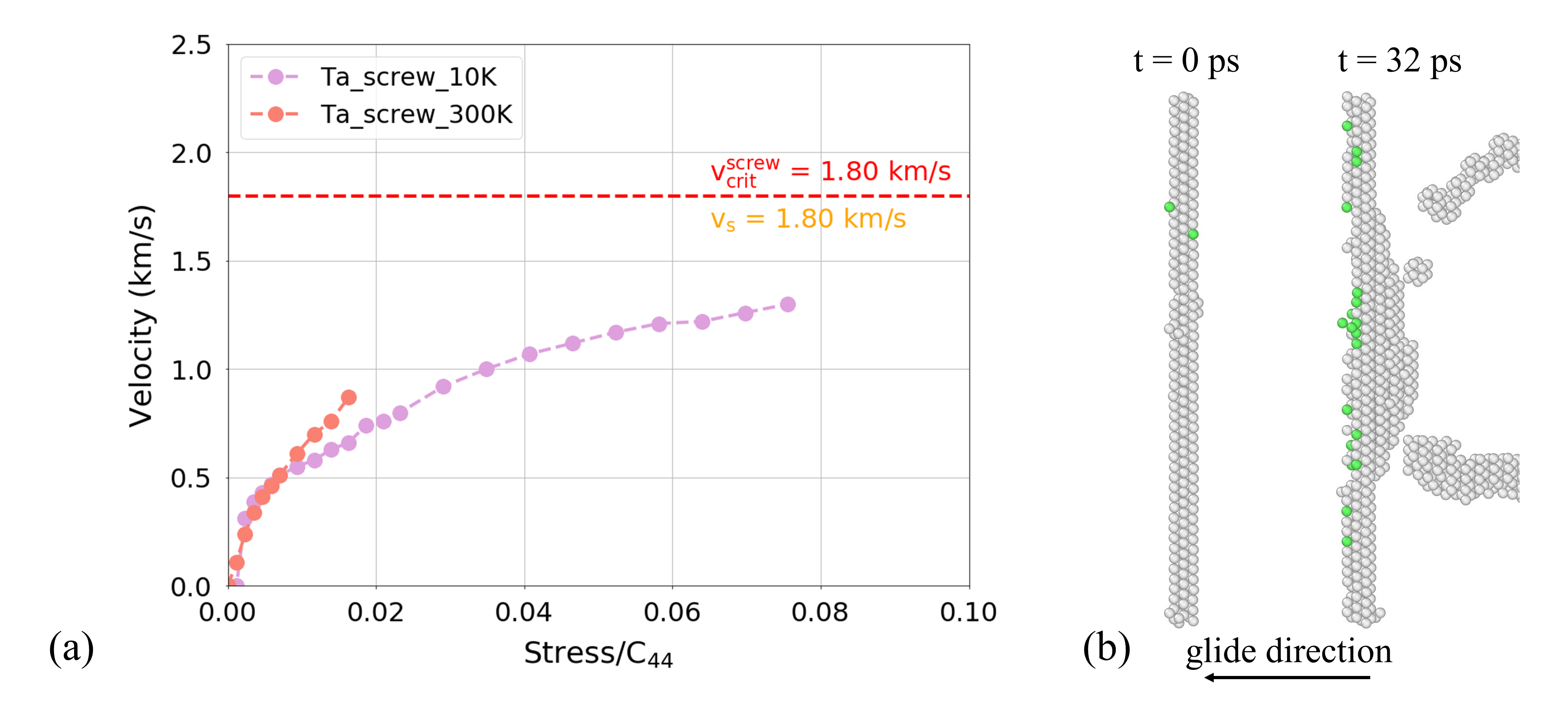}
\caption{We show screw dislocation velocity as a function of applied stress for Ta ($T = 10$ K, $300$ K).
The dislocation was accelerated along the $[0\bar{1}1]$ direction whose lowest shear wave speed for Ta is
$\sim1.80$ km/s (determined from its experimental density and elastic constants in Table \ref{tab:values-metals}) which in this direction coincides with the critical velocity.
}
\label{fig:MD_Ta}
\end{figure}

\begin{figure}[h!t]
\centering
\includegraphics[width=0.60\textwidth]{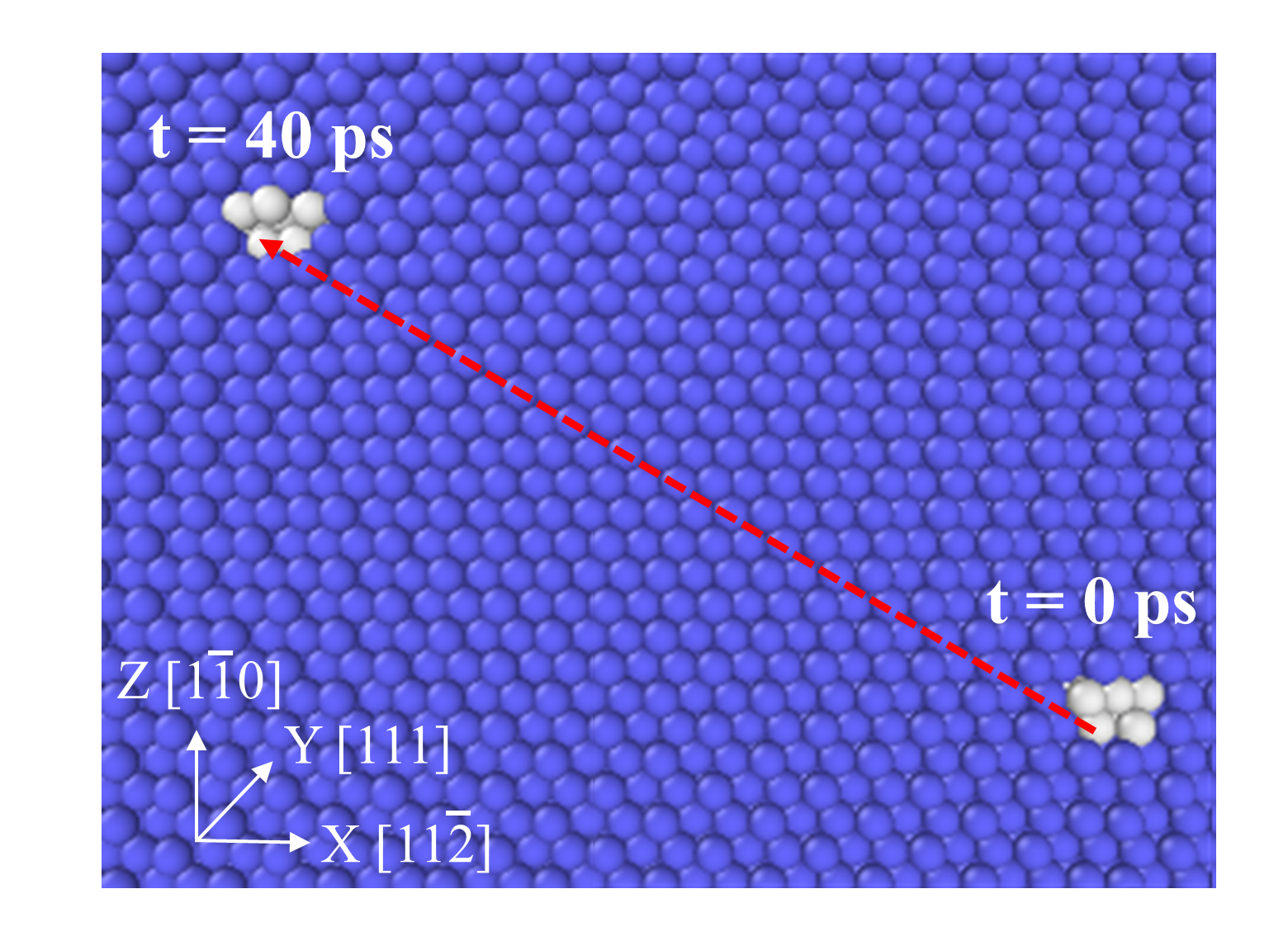}
\caption{Motion of the screw dislocation in Ta under a shear stress of $800$ MPa ($T=10$ K).}
\label{fig:motion_Ta}
\end{figure}

\noindent
Within Table \ref{tab:values-metals} we list the metal densities and SOECs at ambient conditions, as well as the lowest shear wave speeds and critical velocities for screw dislocations determined from those values.
The direction of dislocation motion and sound wave propagation is indicated in brackets next to the shear wave speed values.
Since the values of elastic constants are slightly different within MD simulations, we also list the latter for the metals and temperatures we have simulated.
The shear wave speeds and critical velocities pertaining to our MD simulations remain close to the values determined from experimental SOECs.
All shear wave speeds were calculated from \eqnref{eq:sound1}, the critical velocities for the fcc metals were calculated from \eqref{eq:vcrit_fccscrew}, whereas the critical velocities for bcc Ta were determined numerically from $\det(nn)=0$ within \eqref{eq:ukl-sol} as described in Ref. \cite{Blaschke:2017lten}.

\subsection*{Computational Details}

Fig. \ref{fig:setup} shows the simulation setup.
Screw dislocations are inserted into a simulation cell which is periodic along the line direction and glide direction following the approach discussed in \cite{Maresca:2018}.
The shear stress is applied to the top and bottom layer of atoms along positive $x$ and negative $x$ direction, respectively.
The top and bottom surfaces are free, except along the line direction, in which forces are applied to atoms within 5 \r{A} of each surface to replicate a shear stress, $\tau_{yz}$.
As depicted within Fig. \ref{fig:setup}(a), for Al and Cu the simulation cell size is 30 nm x 30 nm x 20 nm with the glide direction along x, line direction along y and slip-plane normal along z. For the Ta simulation, the simulation cell size is 25 nm x 60 nm x 15 nm, as depicted within Fig. \ref{fig:setup}(b).
We use Mishin's potential for Al \cite{Mishin:1999} and Cu \cite{Mishin:2001}, and Ravelo's potential for Ta \cite{Ravelo:2013}.
The simulations are performed using LAMMPS \cite{Plimpton:1995}, and the results are visualized using Ovito \cite{Stukowski:2009}.

Fig. \ref{fig:MD_Al}(a) shows the variation of the dislocation velocity with applied stress for Al at 10 K and 300 K, and shear wave velocity $v_\txt{s}$ and critical velocity $v_\txt{crit}^\txt{screw}$ are indicated with horizontal dashed lines.
In order to avoid clutter in the plot, we only show the shear wave / critical velocity values calculated from experimentally determined elastic constants, see Table \ref{tab:values-metals}, noting that they are indeed close enough to the values determined explicitly for our MD simulations at 300K and at 0K, so that our conclusions about whether or not dislocations move supersonic in a simulation can be visualized using just the former values.
The calculated velocity here is close to that reported in \cite{Dang:2019}.

The shear stresses are increased up to the point where instabilities appear such that stable dislocation motion cannot be maintained.
For example, when $T = 300$ K, under an applied shear stress of 600 MPa, the screw dislocation in Al develops instabilities in the form of extended dislocation nucleation from the trailing partial, as shown in Fig. \ref{fig:MD_Al}(b).

The results for Cu and Ta are shown in Fig. \ref{fig:MD_Cu}(a) and Fig. \ref{fig:MD_Ta}(a), respectively.
Similar to Fig. \ref{fig:MD_Al}(b), instabilities develop in the form of extended dislocation nucleation at higher stresses (cf. Fig. \ref{fig:MD_Cu}(b) and Fig. \ref{fig:MD_Ta}(b)).
Fig. \ref{fig:motion_Ta} finally shows the gliding of the screw dislocations in Ta at 800 MPa, where the snapshot at 40 ps is superimposed on the initial snapshot.
As marked by the dashed red arrow, the gliding direction in our simulation of bcc Ta is $-\frac12 [11\bar{2}] +  \frac12 [1\bar{1}0]=[0\bar{1}1]$, and the gliding plane is therefore $[0\bar{1}1]  \times [111]=(\bar{2}11)$.

All of our simulations of screw dislocations at room temperature develop instabilities before reaching the critical velocity.
See e.g. Refs. \cite{Tsuzuki:2008,Tsuzuki:2009,Verschueren:2018} and references therein for a general discussion of instabilities in MD simulations.
On the other hand, at very low temperatures (10 K in our case), screw dislocation in Cu can truly overcome $v_\txt{crit}^\txt{screw}$ at shear stresses beyond 2000 MPa (but not at room temperature).


\subsection*{Re-interpreting older MD simulation results in the literature}

References \cite{Olmsted:2005,Marian:2006,Oren:2017} have claimed to observe transonic screw dislocations in Cu and Ni:
In particular
\begin{itemize}
\item Oren et al. \cite{Oren:2017} find stable screw dislocation velocities in Cu which are slightly faster than the shear wave speed (determined by the authors to be $2.15\,$km/s in their simulations).
According to Fig. 5 in that reference, the highest velocity is $2.19\,$km/s and hence between the lowest shear wave speed and the critical velocity (which we have determined to be $2.20\,$km/s but which may be slightly higher in the simulations of Ref. \cite{Oren:2017} since their shear wave speed is slightly higher than ours).
Furthermore, our own simulations suggest that instabilities develop already below the shear wave speed in contrast to  \cite{Oren:2017} (see Figure \ref{fig:MD_Cu}).

\item Peng et al. \cite{Peng:2019} report supersonic screw dislocations in Cu up to 3.5 km/s at very low temperatures (1 Kelvin).

\item Olmsted et al. \cite{Olmsted:2005} found stable screw dislocation velocities in Ni up to $\sim 2.9\,$km/s at 300 Kelvin which is above the lowest shear wave speed ($2.75\,$km/s) at room temperature (see Fig. 9 in that reference).
However, this velocity is still below the true critical velocity of $3.00\,$km/s according to our Table \ref{tab:values-metals}.

That same reference found that screw dislocations in Al exhibit stable motion only well below the lowest shear wave speed
which in turn is below the critical velocity.

\item Marian et al. \cite{Marian:2006} also find stable screw dislocation velocities in Ni up to $\sim 2.9\,$km/s.
Upon inspecting the according Fig. 6 in that reference, we cannot make out any `saturation' or `marked leap into the transonic regime' near the lowest shear wave speed.
Again, the highest velocity found in \cite{Marian:2006} is still below the critical velocity of Table \ref{tab:values-metals}, which we have argued should be categorized as being ``subsonic''.
\end{itemize}

Together with our own MD results, these findings support our conclusion that at room temperature stable screw dislocation motion exists only below the critical velocity which we have argued should be viewed as the velocity separating a ``subsonic'' from a ``transonic'' regime.
Since for fcc metals there is only one critical velocity for screw dislocations, the term ``transonic'' can be used synonymously with ``supersonic'' in this case.
Supersonic screw dislocations only seem possible at very low temperatures.
This view is supported by our present results for screw dislocations in Cu at 10 K shown in Figure \ref{fig:MD_Cu} as well as the results of Ref. \cite{Peng:2019} for Cu at 1 K, which both show true supersonic motion (i.e. above the critical velocity).
The reason Ref. \cite{Peng:2019} reports even higher speeds than we do, is likely that the authors used a very small line length along the Burgers vector (0.767 nm), which might significantly affect the mechanisms of screw dislocation motion.
For example, the kink-pair nucleation mechanism requires greater line length, see \cite{Marian:2004,Domain:2005,Gordon:2010}.
As has been shown in Refs. \cite{Osetsky:2003,Cai:2003,Szajewski:2015a,Oren:2017}, supercell size and the corresponding aspect ratio also affect the dislocation motion, especially at relatively low applied stresses.

In trying to understand why supersonic motion seems only possible at low temperatures we note the following facts:
The derivation of the critical velocities was based on very idealized assumptions, such as a neglected dislocation core and steady-state motion, and as such can not be expected to be a hard limit in real world environments.
In fact, Refs. \cite{Markenscoff:2008,Pellegrini:2018} showed that the divergences can be removed by regularizing the core.
We emphasize though, that a regularized dislocation will not change the value of the critical velocity; rather an expanded core makes it possible, in principle, to overcome a `softer' barrier $v_\txt{crit}$.
At temperatures close to the Debye temperature and higher, phonons scattering on the moving dislocation become an important effect for fast moving dislocations:
Scattered phonons impose a `drag force' (known as phonon wind), and it has been shown \cite{Blaschke:2018anis,Blaschke:2019Bpap} that a divergence in the dislocation field leads to a divergence in the drag force at the critical velocity.
At very low temperatures, only few phonon modes are excited and because phonon drag is most sensitive to high frequency phonons, the drag force is greatly diminished at low temperatures \cite{Alshits:1992}.
We may therefore speculate, that the presence of supersonic screw dislocations at low temperatures and their absence at high temperatures is due to phonon drag making the critical velocity even harder to overcome, and the pure screw dislocations simulated so far become unstable before that happens.
Similarly, lattice waves emitted from the dislocation core as discussed in \cite{Kim:2020,Kim:2020b} may enhance the effect further.
This explanation seems plausible especially for Cu and Ni, where Refs. \cite{Olmsted:2005,Marian:2006} have seen screw dislocations moving at speeds close to the critical velocity.
In other materials, like Al and Ta, thermal fluctuations may be leading to instabilities well below the critical velocity.

\section{Conclusion}

In this paper we have derived analytically (within the theory of linear elasticity) the critical velocity at which the elastic energy of a steady-state screw dislocation in an anisotropic fcc crystal diverges.
This physically derived critical velocity is greater than the lowest shear wave speed in the direction of dislocation motion, a fact that has not been appreciated previously in the literature leading to physically unjustified claims of supersonic screw dislocation motion in fcc metals.

In particular, most existent previous MD simulations \cite{Olmsted:2005,Marian:2006,Oren:2017} to date have found stable steady state motion of fcc screw dislocations below our predicted critical velocity, yet above the lowest shear wave speed in certain cases.
The only exception is Ref. \cite{Peng:2019} reporting screw dislocation speeds well above the critical velocity in Cu at 1 Klevin.
A few references \cite{Olmsted:2005,Marian:2006,Oren:2017,Peng:2019} have interpreted velocities between the lowest shear wave speed and the critical velocity as being `transonic', a view we presently do not share:
As shown via analysis, the lowest shear wave speed
is inconsequential to dislocation dynamics, and rather
the critical velocity separates distinctly different regimes of dislocation motion which we refer to as `subsonic' and `supersonic'.
Since there is only one critical velocity for fcc screw dislocations, there is no transonic regime.

We have also shown results from our own independent MD simulations for fcc Al, Cu, and Ta.
At room temperature, our simulation results seem to corroborate our analytical analyses that no stable supersonic screw dislocations exist, but at very low temperatures, on the other hand, they are in contrast in that they show stable, steady state motion above the critical velocity for Cu at 10 Kelvin.
In other words, at 10 Kelvin we confirm evidence  reported in  Ref. \cite{Peng:2019} of true stable supersonic screw dislocations in Cu (i.e. with velocity $v>v_\txt{crit}$) from MD simulations.
Some references \cite{Oren:2017,Peng:2019} reported higher dislocation speeds in Cu than us.
It is known that not only temperature, but also dislocation line length \cite{Marian:2004,Domain:2005,Gordon:2010} as well as supercell size and aspect ratio \cite{Osetsky:2003,Cai:2003,Szajewski:2015a,Oren:2017} can influence MD results in regard to dislocation instabilities.
It will therefore be worthwhile to further quantify these effects within future work.

The velocity of dislocations under extreme loading conditions is also important in the accurate simulation of high rate plastic deformation. 
First, consider Orowan's relation which relates the product of mobile dislocation density and velocity to the plastic strain rate. 
The existence of an asymptotic subsonic dislocation velocity suggests that in regimes of high applied loadings, changes in the plastic strain rate are governed by the mobile dislocation density.
In regimes of even higher applied loading dislocations may travel supersonically and a lower mobile dislocation density is required to accommodate the same plastic strain rate \cite{Blaschke:2019a}. 
The insights gained in our present work are also applicable to the numerical simulation of dislocations via discrete dislocation dynamics (DDD) methods, e.g. ParaDiS \cite{Arsenlis:2007}.
Within the DDD methodology, prescribed mobility laws relate stress to dislocation velocity. As we have demonstrated, the shear wave speed of a given material differs from the true asymptotic subsonic dislocation velocity. Incorporating true asymptotic subsonic dislocation velocities into various DDD methodologies \cite{Arsenlis:2007,Bertin:2015,Devincre:1997,Ghoniem:2000,Zbib:1998,Cui:2019} will improve simulations in regimes of high strain rate, e.g. $\dot\varepsilon\gg 10^5\,\txt{s}^{-1}$.

\subsection*{Acknowledgements}
\noindent
The authors would like to thank R. G. Hoagland for related discussions.
This work was supported by the Institute for Material Science at Los Alamos National Laboratory.
In particular, D.N.B. and J.C. acknowledge support by the IMS Rapid Response program.

\appendix
\section{Comparing the two anisotropic steady state solutions}
\label{sec:comparison}

\begin{figure}[ht]
\centering
 \includegraphics[width=0.5\textwidth]{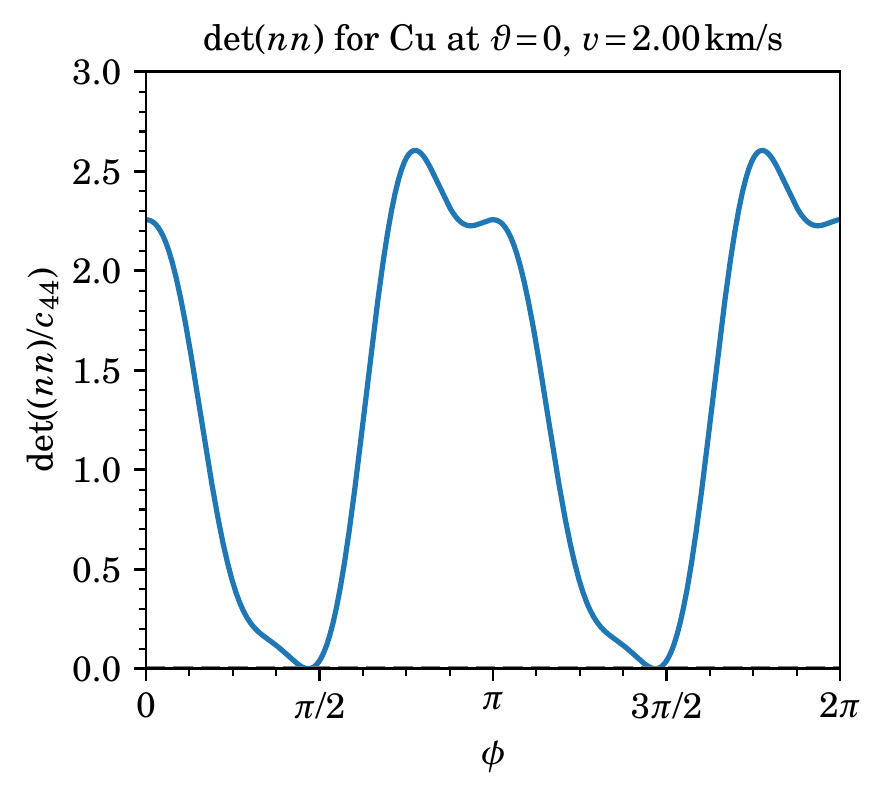}%
 \includegraphics[width=0.5\textwidth]{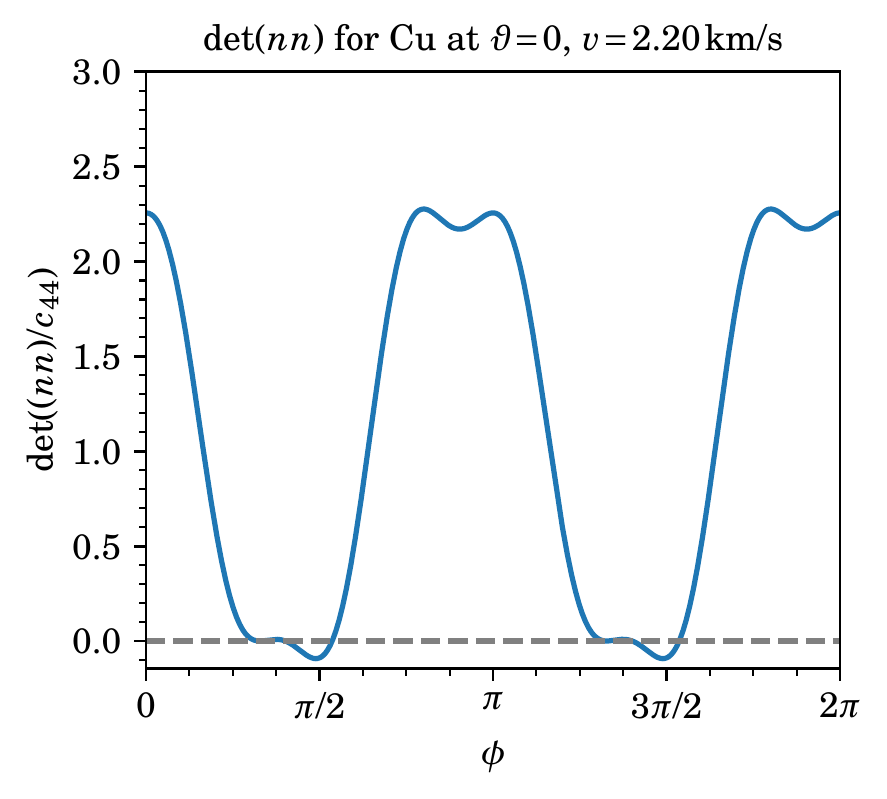}
\caption{We show the determinant $\det(nn)$ for fcc screw dislocations ($\vth=0$) as a function of polar angle $\phi$ at the two critical velocities $\abs{\vec{v}_\txt{c1}}$ (left) and $\abs{\vec{v}_\txt{c2}}$ (right) at the example of Cu where 
$\abs{\vec{v}_\txt{c1}}=2.00\,$km/s, $\abs{\vec{v}_\txt{c2}}=2.20\,$km/s.
(Note that $\abs{\vec{v}_\txt{c1}}$ is slightly lower than the lowest shear wave speed $v_s=2.05\,$km/s, see Table \ref{tab:values-metals}.)}
\label{fig:detNNscrewfcc}
\end{figure}

\begin{figure}[ht]
\centering
 \includegraphics[width=0.5\textwidth]{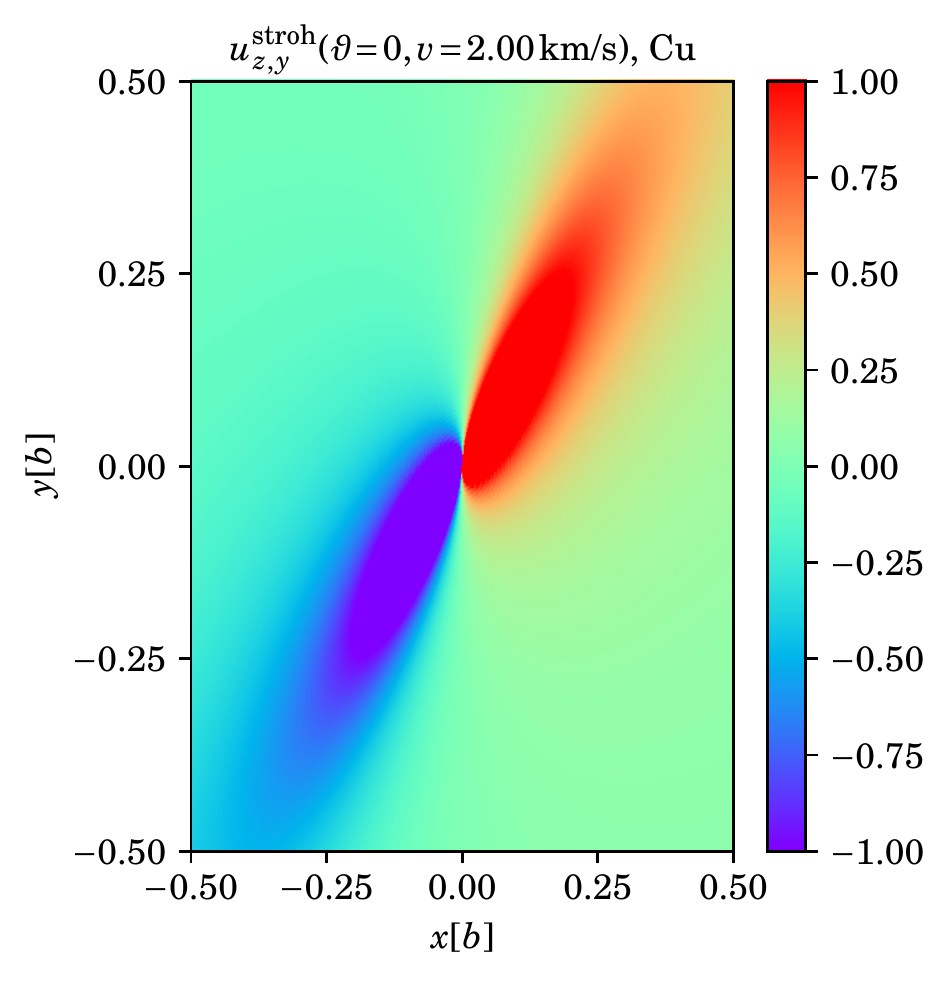}%
 \includegraphics[width=0.5\textwidth]{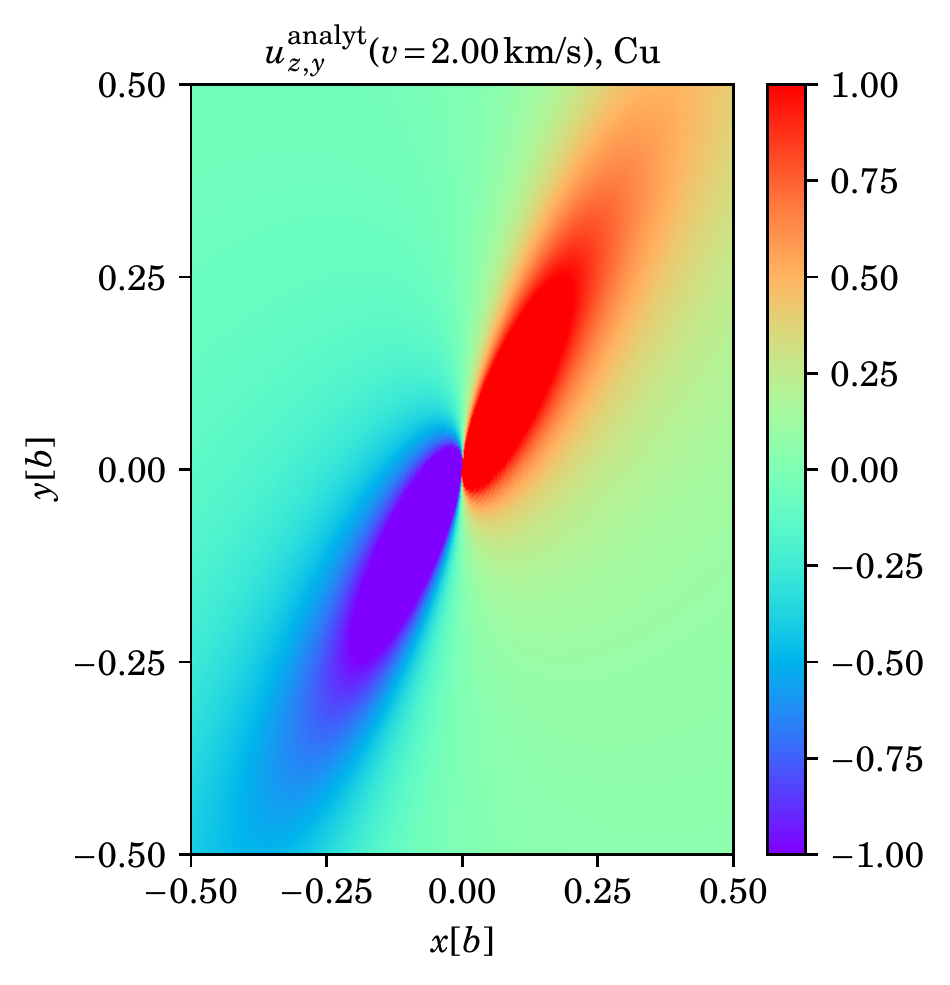}
\caption{We compare $u_{z,y}$ at dislocation velocity
$v=2.00\txt{km/s}\approx v_\txt{c1}$
for Cu of the integral method \eqref{eq:ukl-sol} after rotation into the dislocation frame (left) to the corresponding analytic expression
\eqref{eq:analyticsteadystate_fcc}.
We see that nothing special happens at this velocity and that the two solutions match.
(In the numerical ``integral method'', due to the rounding of $v$ on the one hand and the discretization in $\phi$ on the other hand, we never hit the two discrete points in $(v,\phi)$ space where $\det(nn)=0$ exactly.)}
\label{fig:comparesteadystate}
\end{figure}

\noindent
Let us briefly review the well-known steady state solution using the so-called integral method, a generalization to the Stroh method, see~\cite{Bacon:1980}.
For constant dislocation velocity $v$, the system of equations \eqref{eq:Hooke} can be rewritten as $\hat C_{ijkl}u_{k,il}=0$ with ``effective'' elastic constants $\hat C_{ijkl}\equiv\left(C_{ijkl}-\rho v_iv_l \d_{jk}\right)$.
We have introduced the notation $u_{k,l}\equiv\pa_l u_k$ for the gradient of the displacement field $u_k$.
We further introduce the perpendicular unit vectors $\hat{m}_0(\vth)$ in the direction of dislocation glide and $\hat{n}_0$ as a slip plane normal.
Both of these vectors are also perpendicular to the sense vector $\hat{t}(\vth)$ of the dislocation,
i.e. $\hat{m}_0(\vth) = \hat{n}_0 \times \hat{t}(\vth)$.
Within the integral method, the solution takes the form $u_{j,k}(r,\phi)={\tilde{u}_{j,k}(\phi)}/{r}$ where $\tilde{u}_{j,k}(\phi)$ is a function of the Burgers vector $\vec{b}$, and polar angle $\phi$ contained within the plane of the dislocation line measured with respect to the dislocation cut plane,
\begin{align}
 \tilde{u}_{\!j,k}(\phi)&=\frac{-b_l}{2\pi}\left\{n_k\left[(nn)^{-1}(nm) S\right]_{jl} - m_k S_{jl} + n_k(nn)^{-1}_{ji}K_{il}\right\}\
 \,.\label{eq:ukl-sol}
\end{align}
The vectors $n$, $m$ and tensors $\mat S$, $\mat K$ depend on $\phi$ and the dislocation character angle ($\vth$) through~\cite[p.~476]{Hirth:1982}:
\begin{align}
\vec{m}(\vth,\phi)&=\hat{m}_0(\vth)\cos(\phi) + \hat{n}_0\sin(\phi)
\,,\nn\\
\vec{n}(\vth,\phi)&=\hat{n}_0\cos(\phi) - \hat{m}_0(\vth)\sin(\phi)
\,,\nn\\
\mat S&=-\frac1{2\pi}\int_0^{2\pi}(nn)^{-1}(nm)\,d\phi
 \,, \nn\\
 \mat K&=-\frac1{2\pi}\int_0^{2\pi}\left[(mn)(nn)^{-1}(nm)-(mm)\right]d\phi
 \,,\label{eq:vecs-sol}
\end{align}
with the notation $(ab)_{jk}\equiv a_i \hat C_{ijkl} b_l$.
In order to match sign conventions with respect to the orientation of the dislocation to Section \ref{sec:steadystate_anis} above, we have reversed the overall sign of $\tilde{u}$ within \eqref{eq:ukl-sol} as compared to \cite{Hirth:1982,Blaschke:2017lten}.
Variables $r$, $\phi$ are polar coordinates in the plane spanned by $\hat{m}_0(\vth)$ and $\hat{n}_0$, where $\hat{n}_0$ is the slip plane normal, $\hat{m}_0(\vth)$ is perpendicular to $\hat{n}_0$ and $\hat{t}(\vth) = \frac1b\left[\vec{b}\cos\vth+\vec{b}\times\hat{n}_0\sin\vth\right]$.
Note that $\tilde{u}_{j,k}(\phi)$ includes terms proportional to
$(nn)^{-1}$ and hence exhibits divergences at $\det(nn)=0$.
These divergences occur along a unique path in a plane defined by the polar angle $\phi$ and critical velocity $\abs{\vec{v}_\txt{c}}$.
Since $(nn)$ is a 3x3 matrix, there are in general three solutions (in ascending order) $\abs{\vec{v}_\txt{c1}}$, $\abs{\vec{v}_\txt{c2}}$, and $\abs{\vec{v}_\txt{c3}}$.
In general none of these velocities need to coincide with any particular sound speed, though in practice $\abs{\vec{v}_\txt{c1}}$ is usually very close to or even coincident (cf. fcc edge dislocation) with the lowest shear wave speed propagating parallel to the dislocation motion \cite{Blaschke:2017lten}.

In comparing to our analytic solution for pure screw dislocations \eqref{eq:analyticsteadystate_fcc}, we limit our analyses to $\vth=0$.
Independent of fcc slip system, $\mat{S}\cdot\vec{b}=0$; only the third term of \eqref{eq:ukl-sol} is non-zero.
Interestingly, the smallest solution to $\det(nn)=0$, i.e. $\abs{\vec{v}_\txt{c1}}$ is a ``mild'' one in the sense that at this velocity $\det(nn)=0$ only at two angles $\phi_c$, $\phi_c+\pi$; at those angles $\partial_\phi\det(nn)|_{\phi=\phi_c}=0$.
Figure \ref{fig:detNNscrewfcc} illustrates this nuance at the example of Cu.
Due to the symmetry of the first term in the definition of $\mat{K}$, the integral exists and is indeed \emph{finite} at $\abs{\vec{v}_\txt{c1}}$.
Furthermore, it is readily verified that $\det(nn)(nn)^{-1}\cdot (\mat{K}\cdot\vec{b})$ tends to zero at this velocity leading to a finite limit for $\tilde{u}_{\!j,k}(v=\vec{v}_\txt{c1})$.
The steady state solution (although numerically troublesome at $\vec{v}_\txt{c1}$) exhibits its first divergence at the \emph{second smallest critical velocity} $\vec{v}_\txt{c2}$.
Indeed $\vec{v}_\txt{c2}$
is numerically verified to equal our analytic result
\eqref{eq:vcrit_fccscrew}
shown above in Sec. \ref{sec:steadystate_anis}.
To compare the field solution to its analytic counter part, we have examined both \eqref{eq:ukl-sol} and \eqref{eq:analyticsteadystate_fcc} for five different fcc metals: Ag, Al, Au, Cu, and Ni.
In examining these comparisons, we rotate \eqref{eq:ukl-sol} to a basis coincident with $\hat{x}$, $\hat{y}$, $\hat{z}$ as described within the main text via $\mat{U}^{\txt{fcc}}$, cf. \eqnref{eq:Urot}.
The orientation of $\hat{x}$ (which is parallel or antiparallel to the dislocation velocity) and whether the cut is made at positive or negative $x$ is a matter of convention.
If we rotate such that $\hat{x}$ is parallel to the dislocation velocity, the solution \eqref{eq:ukl-sol} coincides with the gradient of the analytic \eqnref{eq:uz_steadystate_fcc} upon identifying $x=\cos(\phi)$, $y=\sin(\phi)$.
Figure \ref{fig:comparesteadystate} illustrates an example comparison for Cu.
We have verified numerically that $u_{3,1}$ and $u_{3,2}$  coincide to at least 12 significant digits at this velocity and for 1800 angles $\phi\in[0,2\pi]$.

As a final remark, the general steady-state solution \eqref{eq:ukl-sol} is applicable to any crystal symmetry in contrast to \eqref{eq:uz_steadystate_anis}.
Only our discussion on the subtleties of the smallest (would-be) critical velocity are unique for fcc screw dislocations.
Often, such as in fcc edge dislocations or even bcc screw dislocations, the smallest solution to $\det(nn)=0$, i.e. $\abs{\vec{v}_\txt{c1}}$, is indeed the (smallest) critical velocity of the dislocation in question \cite{Blaschke:2017lten}.
In many cases it is very close to (and in the case of fcc edge dislocations coincides with) the lowest shear wave speed corresponding to the direction of dislocation motion.

\bibliographystyle{utphys-custom}
\bibliography{dislocations}

\end{document}